\documentclass[runningheads]{llncs}
\usepackage{graphicx}
\usepackage{enumitem}
\usepackage{tabularx}
\usepackage{comment}
\usepackage[hidelinks,bookmarks,bookmarksdepth=3,bookmarksopen]{hyperref}

\begin{document}
\title{On Collaborative Model-driven Development of Microservices}

%
%
\author{Jonas Sorgalla\inst{1}\orcidID{0000-0002-7532-7767} \and Florian Rademacher\inst{1}\orcidID{0000-0003-0784-9245} \and Sabine Sachweh\inst{1} \and Albert Z\"undorf\inst{2}}
\authorrunning{J. Sorgalla et al.}
%
\institute{Institute for the Digital Transformation of Application and Living Domains\\
	University of Applied Sciences and Arts Dortmund\\
	Otto-Hahn-Stra\ss{}e 23, 44227 Dortmund, Germany\\
	\email{\{jonas.sorgalla,florian.rademacher,sabine.sachweh\}@fh-dortmund.de}
	\and
	University of Kassel, Department of Computer Science and Electrical Engineering\\
	Software Engineering Research Group\\
	Wilhelmsh\"oher Allee 73, 34121 Kassel, Germany\\
	\email{zuendorf@uni-kassel.de}
}
\maketitle 
\begin{abstract}
Microservice Architecture (MSA) denotes an emerging architectural style for distributed and service-based systems whereby each microservice is highly cohesive and implements a single business capability. A microservice system consists of multiple, loosely coupled microservices. It provides complex capabilities through services interacting in choreographies. A single dedicated team, typically practicing DevOps, is responsible for each microservice, i.e., it ``owns'' the service. However, while systems relying on MSA have several architectural advantages especially for cloud applications, their realization is characterized by an increased accidental complexity due to redundant handcrafting of implementation, e.g., to make each service standalone runnable. A promising way to cope with such complexity is the usage of Model-driven Development (MDD) whereby models are used as first-class entities in the software development process. Although there are already first steps taken on how MDD could be applied by a single team to implement its microservices, the question of how MDD can be adapted to MSA's development distribution across multiple teams remains an issue. In this paper we envision the application of Collaborative Model-driven Software Engineering (CMDSE) to MDD of MSA by surveying relevant characteristics of CMDSE and identifying challenges for its application to MSA. The present paper takes a first step towards enabling holistic MDD of MSA across microservice teams.

\end{abstract}
\section{Introduction and Background}
Microservice Architecture (MSA) denotes an emerging architectural style for distributed and service-based systems \cite{Francesco.2017}. As such, MSA relies on the \textit{service} concept as the fundamental architectural building block for a system's architecture. Each microservice is highly cohesive and represents a single business  capability. Technically, a microservice is realized as an independent process that can be managed, i.e., designed, developed, deployed, and operated, autonomously. To realize complex business capabilities, multiple of these services can collaborate in service choreographies through interfaces \cite{Rademacher.2017}. Hereby, the service interaction is generally stateless and uses protocols like HTTP or AMQP\footnote{\url{https://www.amqp.org}} \cite{Newman:2015:BM:2904388}. Furthermore, each microservice is organizationally aligned to exactly one service team which usually practices DevOps \cite{Kang2016_DevOps}. Resulting applications relying on MSA are, among other characteristics, vertical as well as horizontal scalable, flexibly extensible and have short release cycles which makes them especially suitable for cloud applications like Spotify or Netflix \cite{Newman:2015:BM:2904388}. 

However, the advantages of MSA in terms of increased scalability, resilience and technology heterogeneity \cite{Newman:2015:BM:2904388} come at the cost of an increased accidental complexity regarding the overall system development \cite{Rademacher.2017}. One reason for this increased complexity is that microservice architectures, compared to monolithic applications, are distributed by nature \cite{France2007}. Resulting from this distribution, the realization of multiple services involves extensive and redundant handcrafting of implementation, e.g., to make each service independently runnable, or to provide and consume the necessary interfaces for complex operations \cite{Wizenty2017}. 

An approach to cope with the \textit{accidental complexity} of complex, distributed software systems such as MSA is Model-driven Development (MDD) \cite{France2007}. MDD denotes the usage of models as first-class entities in the software development process. Applied to MSA, developers would use a modeling language to design services and use a Model-to-Code (M2C) transformation to (semi-)automatically derive service code \cite{Rademacher.2018MADE}. In such a model-centric development scenario, modeling does not completely replace programming. Instead, the usage of models aims to ease accidental complexity by helping to avoid redundant programming, but does not replace the manual realization of \textit{essential complexity}, e.g., programming service-specific, business-related behavior \cite{Schmidt.2006}.    

Although there are first approaches, e.g., \cite{Dullmann:2017:MGM:3053600.3053627} or \cite{Sorgalla.2017}, which address such an MDD for MSA (MSA-MDD), they currently only enable the generation of a microservice landscape from a centralized architectural perspective. Hence, we argue that a holistic approach to MSA-MDD needs to take MSAs organizational characteristics into account. That is, like the code-centric development process, a model-centric development process of MSA would need to consider Conway's Law in the context of MSA \cite{Newman:2015:BM:2904388} and support a collaborative development spread across multiple teams \cite{Sorgalla.2018}. 

In this paper, we present our vision of a collaborative modeling approach for MSA. For this purpose, we rely on methods and techniques from the research area of Collaborative Model-driven Software Engineering (CMDSE) \cite{DiRuscio.2017}. It defines approaches where multiple stakeholders use a set of shared models to collaborate.

The remainder of this paper is organized as follows. In Section \ref{sec:scope} we elaborate on the collaborative aspects of microservice development and deduce challenges for a corresponding holistic MSA-MDD approach. Building on this, in Section \ref{sec:vision} we describe our vision of a collaborative MSA-MDD approach by applying concepts from the area of CMDSE. Finally, Section \ref{sec:conclusion} concludes the paper and Section \ref{sec:futurework} describes future work. 

\section{Challenges for Collaborative Model-driven Microservice Development}\label{sec:scope}
In this section we identify and discuss major challenges for collaborative modeling in the context of MSA to enable holistic MSA-MDD. CMDSE is itself part of the broader research area of Collaborative Software Engineering (CoSE) \cite{CoSE.2010}, which investigates means for enhancing collaboration, communication and coordination (3C) among software engineers and project stakeholders. In the context of CoSE, the organizational structure of microservices can be separated into two hierarchical scopes of collaboration. In the \textit{team-internal scope}, team members collaborate to manage one or more services. In the \textit{team-external scope}, teams themselves collaborate with each other, e.g., by using an interface of another team's service for their service's realization. Furthermore, the act of assembling the overall system through autonomous services in its own right represents a form of team-external collaboration. 

A holistic MSA-MDD may then be enabled by applying CMDSE to both team scopes. Based on 3C, full-fledged CMDSE approaches comprise the three main complementary dimensions \textit{model management}, \textit{collaboration means}, and \textit{communication means} \cite{TSE.2017}. Each of the following subsections identifies and discusses challenges for collaborative MSA-MDD by analyzing the team-internal scope (cf. Subsection~\ref{sub:team-internal}) and the team-external scope (cf. Subsection~\ref{sub:team-external}) with respect to these three dimensions of CMDSE.

\subsection{Team-internal Model-driven Microservice Development}\label{sub:team-internal} 
A team, which is responsible for one or more microservices, follows a \textit{share-nothing} philosophy to foster agility and autonomy \cite{Francesco.2017}. Therefore, each team is independent from other teams and services in their choices related to services' implementation regarding, e.g., programming languages, databases or employed tools. For example, this autonomy enables a single team to adopt an MDD approach for their services even if other teams do not use MDD \cite{Rademacher.2018MADE}. 

However, in practice the team's technology stack and development process model is often influenced by an organization's culture \cite{Nadareishvili:2016:MAA:3002814}, e.g., if the usage of GitLab\footnote{\url{https://www.gitlab.com}} for managing the software lifecycle has proven successful in existing teams, a new team is highly likely to adopt GitLab, too. In certain cases, choices can be predetermined by the overall organization in order to maintain compatibility, e.g., with an existing deployment pipeline \cite{Balalaie.2016}.

At this level, the possible application of MSA-MDD only differentiates itself from a traditional model-driven development process through the different roles within the  team \cite{Brambilla.2012}. To utilize collaboration across team members, existing solutions, e.g., emfCollab\footnote{\url{http://qgears.com/products/emfcollab}} or the Eclipse Dawn Project\footnote{\url{https://wiki.eclipse.org/Dawn}} can be applied. Such solutions already realize means for the CMDSE dimensions model management and collaboration \cite{TSE.2017}. Depending on the tool, separate communication means like an instant messenger could be added to the collaboration tool stack. However, while these tools provide good means for collaboration, they still need an underlying modeling language for the microservice domain \cite{Rademacher.2018}. This motivates the first challenge for a collaborative MSA-MDD approach:

\paragraph{(C1) Support for Role-specific Team Tasks}
For a model-centric development, this modeling language needs to support the different tasks and roles inside a DevOps-based MSA development team, i.e., the complete management process of a microservice.

\subsection{Team-external Model-driven Microservice Development}\label{sub:team-external}
While the team-internal collaborative modeling scope can be covered leveraging existing CMDSE approaches (cf. Subsection~\ref{sub:team-internal}), especially for the application of collaborative MSA-MDD with regards to the team-external scope MSA-specific challenges arise which we discuss in the following.

With the distribution and loose coupling of functionality and service teams, MSA might not exhibit a central architecture viewpoint or entity, e.g., a team of dedicated software architects, for the overall microservice landscape. However, we expect that such a viewpoint or entity can be of great benefit in the context of MSA. First, it may be aware of the overall team structure and foster communication \cite{Rademacher.2018}. Second, it may document and comprehend the overall static structure and service interactions of a microservice architecture. Models are predestined to represent such structures and interaction relations \cite{Combemale:684289}. Therefore, the second challenge for a collaborative MSA-MDD approach arise:  

\paragraph{(C2) System Model Assembly Across Autonomous Microservices}
How can such a holistic and model-based overview of an MSA be assembled from the involved services and interactions in a loosely coupled way, i.e., without contradicting MSA's paradigm of autonomous services.
\\\\
Another aspect regarding the team-external scope results from Conway's Law. Due to the loose coupling of microservices, the responsible teams also collaborate more loosely preserving their autonomy \cite{Nadareishvili:2016:MAA:3002814}. Although there are mechanics to provide knowledge exchange across teams, e.g., Spotify joins persons with a similar skill set from different teams to horizontal organization structures called \textit{guilds} \cite{Kniberg2012}, knowledge exchanges generally happen on a non-technical and informal level \cite{Wiedemann.2017}. However, next to provided interfaces of other teams' services, teams may also access the source code of microservices, e.g., through company-wide available code repositories or verbal requests \cite{Nadareishvili:2016:MAA:3002814}. This agile opportunities also need consideration in a collaborative MSA-MDD approach:

\paragraph{(C3) Collaboration Means for Teams}
Like source-code, the models of a team need to be accessible and usable for other teams, e.g., to copy domain concepts \cite{Rademacher.2018} or retrieve interface descriptions, without contradicting the loose coupling characteristic of services and teams.  

\section{A Collaborative Modeling Approach for Model-driven Microservice Development}\label{sec:vision}
\begin{figure}
	\includegraphics[width=\textwidth]{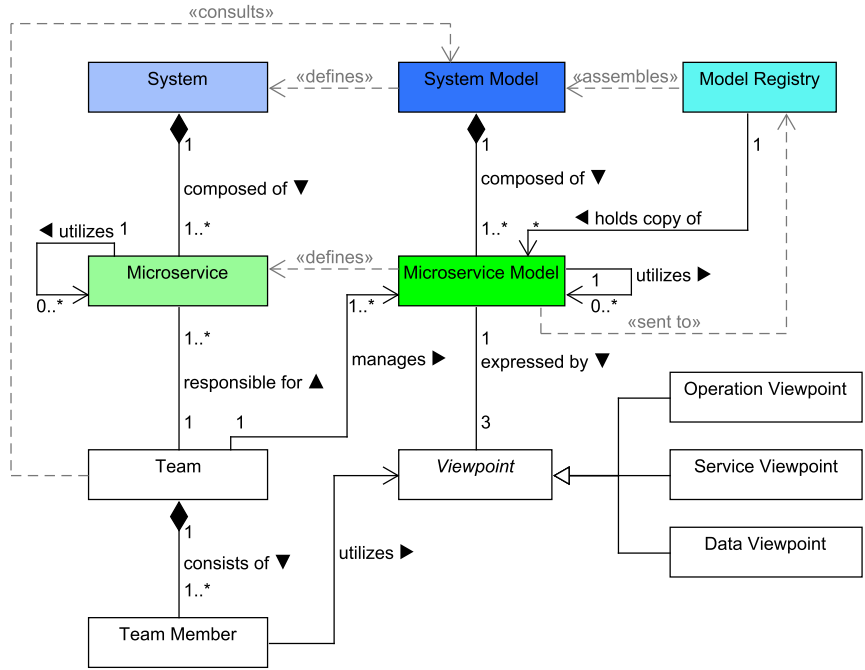}
	\caption{Conceptual Model of Collaborative MSA-MDD} \label{fig1}
\end{figure}

Starting from the identified challenges and their discussion in Section \ref{sec:scope}, we derived a conceptual model for the prospective application of CMDSE to MSA. It is depicted in Figure~\ref{fig1} as a UML class diagram enriched by indirect use relations. 

The overall microservice \texttt{System} is composed of many \texttt{Microservices}. For each of these services, a single \texttt{Team} which consists of multiple \texttt{Team Members} is considered responsible. For a model-centric development, our approach comprises a dedicated \texttt{Microservice~Model}. Next to its services, each team is thus also responsible the defining models \cite{Rademacher.2018}. For the  model-centric development inside a team, we suggest the usage of a separate model repository for each microservice model as means of management and a web- or eclipse-based workbench which works with a local model version repository.

With regard to the role-specific development tasks (cf. C1 in Subsection~\ref{sub:team-internal}), we propose the usage of such a model repository in combination with a set of integrable domain-specific modeling languages (DSMLs), which each addresses a specialized viewpoint for microservice development \cite{RademacherECSApre.2018}. The common metamodel of the DSMLs defines three viewpoints. First, the \texttt{Data~Viewpoint} holds concepts to specify a microservice’s information model. Second, the \texttt{Service~Viewpoint} provides means to model interfaces and dependencies to other teams' services. Third, the \texttt{Operation~Viewpoint} enables team members to model the information for deployment and operation of a service. 

To compose an overall \texttt{System~Model} (C2), our conceptual model involves a central \texttt{Model~Registry} and an additional step in the continuous delivery pipeline, when a team releases a microservice. In this step, a copy of the corresponding service model is sent to the central model registry every time a microservice gets released. Thus, the registry is able to assemble the system model by weaving the microservice models according to their interface dependencies with other services. To ensure a successful composition of the system model, each microservice model is tested at each release for its integrability. A model which integration test fails, e.g., because an external service refers to a data object that is no longer published through the service's interface, is therefore marked as a conflict and has to be revised by the respective team.     

As a result, our presented approach is able to provide teams with the ability to consult other teams' models through the assembled system model (C3). For realization, we envision the extension of the team-internal modeling workbench with the ability to access the system model and import other microservice models as dependencies inside the teams own model. While dependency information gets pushed to the model repository in the next release, the system model can also be consulted regarding change impact and conflict analysis \cite{1235428}, and perform appropriate measures, e.g., automatically protection of deprecated microservice releases because of other services' dependencies. 

\section{Conclusion}
\label{sec:conclusion}

The usage of MDD for designing MSA is a promising way to cope with MSA's inherent accidental complexity. While there already exist approaches for MSA-MDD which support the development from a central architectural point of view,  MSA's organizational characteristic of aligning services to teams is currently underrepresented. 

Hence, we identified three major challenges for the realization of a holistic MSA-MDD across microservice teams by examining team-internal and team-external collaboration processes in microservice development (cf. Section \ref{sec:scope}).

As a result, we presented our vision of a collaborative MSA-MDD approach which foresees individual microservice models as model fragments of the overall system  (cf. Section \ref{sec:vision}). Leveraging a model registry, such models get automatically woven to a system model which in the following can be used to provide team collaboration means, e.g., partial imports or dependencies of other microservice models across teams.

\section{Future Work}
\label{sec:futurework}
For future work we are going to evaluate existing CDMSE approaches like Mondo\footnote{\url{http://www.mondo-project.org}} or Eclipse Dawn\footnote{\url{https://wiki.eclipse.org/Dawn}} for their applicability towards team-internal collaboration and extendability concerning our envisioned approach. In the following we plan to adapt our central modeling approach described in \cite{RademacherECSApre.2018} to support a distributed modeling and implement a prototype for the model registry mechanism.

Beyond the realization of a model-centric development process at design time, we would like to further investigate the possibilities of runtime models \cite{France2007} in the MSA software life cycle. Another interesting research direction we would like to further investigate comprises the usage of microservices as containers for language components in the context of globalizing modeling languages \cite{combemale:hal-00994551}. 
%
%
%
%
\bibliographystyle{splncs04}
\bibliography{literature}
\end{document}